# Robust and Efficient Sorting with Offset-Value Coding

THANH DO*, Celonis Inc
GOETZ GRAEFE, Google Inc

Sorting and searching are large parts of database query processing, e.g., in the forms of index creation, index maintenance, and index lookup; and comparing pairs of keys is a substantial part of the effort in sorting and searching. We have worked on simple, efficient implementations of decades-old, neglected, effective techniques for fast comparisons and fast sorting, in particular offset-value coding. In the process, we happened upon its mutually beneficial relationship with prefix truncation in run files as well as the duality of compression techniques in row- and column-format storage structures, namely prefix truncation and run-length encoding of leading key columns. We also found a beneficial relationship with consumers of sorted streams, e.g., merging parallel streams, in-stream aggregation, and merge join. We report on our implementation in the context of Google's Napa and F1 Query systems as well as an experimental evaluation of performance and scalability.

## 1 INTRODUCTION

Sorting is a fundamental operation for database query processing. Let three examples suffice here: First, in preparation of index nested loops join, efficient index creation requires sorting index entries. Second, in-stream aggregation and merge join require sorting on join or grouping columns, with external merge sort enabling early aggregation while merging runs [3] or while forming initial runs using an in-memory index. Third, hash aggregation and hash join require partitioning, i.e., a distribution sort on hash values.

Our immediate need for efficient external merge sort and for fast comparisons arose in the context of data warehousing and relational query processing using Google's Napa [1] and F1 Query [25, 27] systems, with data arrival rates of many GB/sec and multiple EB/day. External merge sort is required for index creation, e.g., b-trees [2, 6, 15], sort-based query execution, e.g., join and "top" algorithms, and stream indexing, e.g., log-structured merge-trees [24] and stepped-merge indexes [18]. The prevalence of sorted storage structures in our data centers strongly suggests that our query processing software should exploit interesting orderings [26].

Knuth [19] reported estimates from the 1960s that 25% or even 50% of all computer time was spent on sorting and searching. Goetz [11] wrote that in 1960, "because sorting often consumed 50% or more of an application's processing time, there was an intense need for better and faster sorts." Our experience with relational query processing suggests that those estimates might still apply half a century later. Therefore, we explored techniques to speed up external merge sort.

After some benchmarking and profiling, our initial efforts focused on reducing the count and the cost of comparisons within external merge sort. We eventually realized that offset-value [7] coding offers many benefits heretofore not mentioned in the literature: there is a strong link from offset-value coding to compression by prefix truncation and to compression by run-length encoding; and offset-value coding speeds up comparisons not only within sorting but also in other sort-based algorithms for database query processing, e.g., merge join as well as in-stream grouping and aggregation.

**111**

---

Authors' addresses: Thanh Do, T.Do@Celonis.com, Celonis Inc; Goetz Graefe, Google Inc, GoetzG@Google.com.

Table 1. Key values and offset-value code after a comparison.

| Column values | | | | Offset | Value | Offset-value code (after this comparison) |
|---|---|---|---|---|---|---|
| A | B | C | D | | | |
| 11 | 22 | 33 | 43 | | | |
| 11 | 22 | 33 | 44 | 3 | 44 | $400 - 3 \times 100 + 44 = 144$ |

External merge sort can perform its comparisons very efficiently by turning most of them into inline comparisons of integer values. It requires normalized keys and their fixed-size prefix, also known as poor man's normalized keys. A dynamic form of poor man's normalized keys is offset-value coding [7], which combines prefix truncation, inline execution of comparisons in priority queues and merge logic, poor man's normalized keys, and (in some architectures) hardware support [17].

A row's offset-value code is always relative to another row earlier in the sort order: the offset indicates the first difference between them (the first column or the first byte) and the value indicates the data value at that offset, suitably normalized for ascending vs descending sort, etc. Offset and value are combined into a single order-preserving integer. If two rows' offset-value codes are relative to the same base row, different offset-value codes decide the row comparison and render comparisons of column values obsolete.

Table 1 illustrates two four-column key values and their encoding after a comparison. Only the high key value, i.e., later in the sort order, is encoded after the comparison. The encoding is relative to the lower key value and reflects completed comparison effort. Offset-value coding applies to either columns or bytes; here, four columns with domain 1..100 are encoded. The value 400 is chosen in this example to ensure that all offset-value codes are non-negative numbers. The offset is multiplied with the domain size and subtracted because a large offset indicates a lower key value, i.e., earlier in the sort order. Other ways of calculating offset-value codes are also possible [17].

To our surprise, we found a tight relationship (not previously discussed in the literature) between efficient comparisons and compression, i.e., offset-value coding in the sort logic and run sizes on storage. Compression here applies to both intermediate runs on temporary storage and to the final sort results, in the forms of both prefix truncation in row-formats and run-length encoding in column-formats. Moreover, inline execution of preliminary comparisons can reduce overall comparison effort, quite similar to comparisons of hash values in hash-based query execution. Our experiments show, compared to a traditional efficient implementation of external merge sort, 10 to 20 times less effort for key comparisons as well as 10-50% less space, copying, and I/O for intermediate sort results. We also found opportunities to exploit offset-value coding in query execution and data processing beyond sorting.

Among the following sections, Section 2 describes our application context of data warehousing, log-structured merge-trees, and delta compaction. Section 3 reviews related prior work as well as the principal techniques for external merge sort, priority queues, and efficient comparisons. Section 4 describes the techniques we implemented in order to speed up comparisons and reduce efforts for copying and for I/O. Section 5 reports on successes and failures based on measurements and execution profiles. The final section summarizes and concludes.

## 2 APPLICATION CONTEXT: QUERY PROCESSING IN DATA WAREHOUSES

The application context that motivated this work is relational query processing over large and complex data warehouses. The tables and indexes in our application context are huge: often many billions of rows and many tens of columns, including array-valued columns. Large keys are also common, e.g., 20 or even 50 columns, with few key values, e.g., 3 to 10 distinct values per column. Therefore, sorting often requires deep comparisons, e.g., 10 or even 20 columns. For example, a table with 10B rows and key columns with 3 distinct key values requires more than 20 key columns in a unique key [$10^{10} \approx 3^{21}$ or $\log_3(10^{10}) \approx 21$].

Some of the processing is delegated to specialized dataflow environments such as Map-Reduce [9] or Flume [4]. This includes creation and maintenance of materialized views and their storage structures [1]. These are usually sorted, e.g., as b-trees [2, 6, 15]. Queries over these storage structures use a query processing engine that is relational at its core [25, 27].

Sorting is used for many purposes in this context. Even a simple analysis such as counting unique visitors per day or month requires grouping billions of log records, and many analyses require joining a stream of log records with stored information.

Whether these analyses are implemented in a paradigm like Map-Reduce [9], in a dataflow engine such as Flume [4], or as relational query execution plans [13], there are three approaches to processing and matching large datasets. First, persistent indexes such as b-trees can be created, maintained, and searched in algorithms such as index nested-loops join or look-up join. Second, data sets can be sorted and processed by simple, efficient algorithms such as in-stream aggregation and merge join. Third, in-memory hash tables are faster than persistent indexes; beyond hash tables, partitioning to disjoint overflow files copes with very large inputs or, more generally, with steps in a memory hierarchy.

Note that index creation as well as index maintenance benefit from sorting future index entries. This is true not only for ordered indexes such as b-trees but also for spatial indexes [22]. Efficiency of creation and maintenance as well as operational capabilities such as phantom protection by key-range locking strongly suggest b-trees on hash values as an implementation technique for database hash indexes.

Note also that hash-based query execution algorithms, in particular hash partitioning to overflow files, are special cases of distribution sort [16]. Hash values as sort keys not only enable fast comparisons but also promise balanced partitioning with short, uniformly distributed key values. Taken together, whether a dataflow engine or query execution engine relies on indexes, on traditional sorting, or on hash-based query execution algorithms, the core ingredient that permits efficient algorithms for large data sets is sorting.

In our context, sorting and in particular merging are the core algorithms in ingesting, indexing, and querying streams. Log-structured merge-trees and stepped-merge indexes [18, 24], as the names suggest, principally rely on merging and use merge steps akin to external merge sort.

## 3 RELATED PRIOR WORK

External merge sort is, of course, an old, proven, and well-known technique. This section reviews some related techniques either less well known or required for understanding the new techniques introduced in a later section.

### 3.1 External merge sort

External merge sort has three phases: run generation producing initial sorted runs on temporary storage, intermediate merge steps producing intermediate runs, and a final merge producing output. If the sort operation is embedded within a query execution plan [14, 20], the input phase competes for resources such as memory with operations producing the input and the output phase competes with operations consuming the output and with concurrent operations, e.g., two sort operations producing the two inputs into a merge join.

Run generation in read-sort-write cycles, typically using quicksort as the internal sort algorithm, produces runs the size of the memory dedicated to the sort operation. Quicksort on average invokes 10-30% more comparisons than the theoretical minimum of $\log_2(M!) \approx M \times \log_2(M/e)$ for $M$ rows in memory and Euler's constant $e \approx 2.718 \approx 19/7$.

Alternatively, sorting a memory load of $M$ rows can be implemented as merging M runs with a single row each. A tree-of-losers priority queue (see Section 3.2 below) can hold the count of comparisons very close to the theoretical minimum as each input row participates in precisely $\log_2(M)$ comparisons during run generation with $M$ rows in memory.

Run generation using replacement selection, typically using a priority queue, can also produce longer runs. For random inputs, the expected run size can be 2× the memory size. In the best case, i.e., when the input is sorted or nearly so, the entire input forms in a single run equal to the final output. In the worst case, i.e., when the input is sorted in reverse order, each run size equals the memory size, i.e., the run size in read-sort-write cycles.

This technique can also be implemented as merging $M$ runs. After a row is evicted to the current initial run on temporary storage, it is immediately replaced by a row from the unsorted input. The key values of these two rows are compared and the new input row is considered a continuation of the merge input run if the new key value is larger than the evicted key value, i.e., later in the desired sort order. Otherwise, the new input row is assigned to the next initial run on temporary storage. Note that the additional comparison doubles the expected run size, which reduces the number of comparisons during merging by one for each row. In other words, the expected number of comparisons per row or per sort operation remains unchanged.

Some sort operations extend a pre-existing list of sort keys. For example, an input might be sorted on columns A and B but the output is needed sorted on columns A, B, C, and D. In this case, it is sufficient to sort input rows on columns C and D for each distinct A-B pair. A sort in such "segmented" execution competes for resources with both its producing and consuming operations.

## 3.2 Tree-of-losers priority queues

Priority queues are in-memory data structures, typically using a binary tree mapped to an array such that the tree node in array slot $j$ has its parent in array slot $j/2$; the tree's root node is in array slot 1. The principal invariant in most priority queue implementations is that the key value in a parent node is lower (in an ascending sort) than the key values in both (or all) of its children. When a priority queue removes the lowest key value from the root node, a replacing key value moves from the root towards the tree's leaf level with two comparisons per level, one comparison to determine the lower key value among a parent's two children and another comparison to determine whether this child and the parent should switch node contents.

A tree-of-losers priority queue also uses a binary tree embedded in an array, with the tree's root node in array slot 0. The source of its efficiency is its reliance on leaf-to-root passes with one comparison per tree level. Root-to-leaf passes with two comparisons per tree level are not required.

Figure 1, copied from [19], shows a tree-of-losers priority queue. Each rectangular node represents a run, e.g., in a merge step of an external merge sort. Round nodes are tree nodes, with the array's slot number outside and the key value inside the circle. The principal invariants are that two candidate key values compete at each node and that after a comparison of these key values, the loser remains in the node and the winner becomes a candidate in the next tree level. Thus, the overall lowest key value reaches the root node in array slot 0.

When merging runs, two runs are assigned to each leaf node, with one key value remaining there and one key value moving up. When writing the overall winning key value to the merge output, the successor key values from the winner's original input enters at the assigned leaf node and a complete leaf-to-root pass moves the next overall winner to the root again.

While the traditional priority queues can grow and shrink quite easily, tree-of-losers priority queues require either a clever assignment of runs to leaf nodes or complex restructuring when growing or shrinking. Thus, end-of-stream while merging runs usually is managed using an artificial fence key value $+\infty$. A fence key value of $-\infty$ can also be useful, e.g., during initialization of a tree-of-losers priority queue.

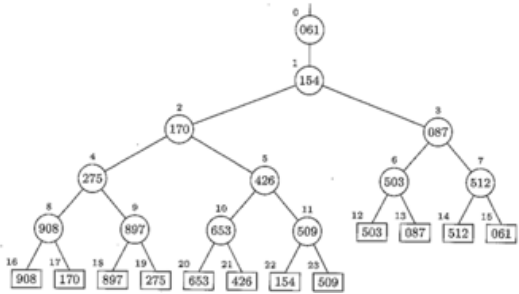

Fig. 1. A tree-of-losers priority queue.

With only leaf-to-root passes, run generation and merging with tree-of-losers priority queues can guarantee sort operations with near-optimal comparison counts, i.e., very close to $\log_2(N!) \approx N \times \log_2(N/e)$ for $N$ rows. With good expected and worst-case complexity yet very limited implementation complexity, it is no surprise that some architectures include dedicated hardware support for tree-of-losers priority queues, e.g., the UPT "update tree" instruction on IBM's mainframes [17].

### 3.3 Normalized keys

Real-world sort keys can be complex due to multiple columns with different sizes and types, null values, compression and encoding, and local alphabets and sorting rules. In order to simplify and speed up comparisons, a common technique maps each row's entire key to a single binary string such that a binary comparison of such strings is order-preserving with respect to all sorting rules. Even if this key normalization is more complex than a single comparison, it is done only once for each row and speeds up $\log_2(N)$ comparisons (assuming $N$ input rows).

For additional speedup, or as an alternative to full normalized keys, the first few bytes of a normalized key can form an integer value, called a poor man's normalized key. The sort logic may include inline comparisons of such integers [23], comparing entire binary strings only for comparisons of two rows with equal poor man's normalized keys. While hardware-supported comparisons of binary strings are faster than column-by-column comparisons with interpretation of column type, size, and sort order (international locale, ascending/descending, null values, etc.), a single integer comparison compiled into the sort code is even faster.

A poor man's normalized key may capture more than just the key value. For example, if a priority queue implementation uses artificial fence key values $-\infty$ and $+\infty$, they may be the lowest and highest values of a poor man's normalized key. If a priority queue manages key values assigned to two runs, the run number may also be part of the poor man's normalized key.

Thus, comparisons of poor man's normalized keys are effectively free, because they cost no more than a test whether a key value is valid or not, e.g., due to end-of-stream of one of the merge inputs.

### 3.4 Offset-value coding

When the sort key has many columns, a poor man's normalized key can cover only a limited number of leading key columns. If the leading columns have few distinct values, the poor man's normalized key may decide only a small fraction of a sort operation's comparisons. In those cases, a dynamic poor man's normalized key can adjust which key columns it represents. This is called offset-value coding [7].

Table 2. Alternative offset-value code calculation after a comparison.

| Key values | | | | Offset | Value | Offset-value code (after this comparison) |
|---|---|---|---|---|---|---|
| A | B | C | D | | | |
| 11 | 22 | 33 | 43 | | | |
| 11 | 22 | 33 | 44 | 3 | 44 | $(1 + 3) \times 100 - 44 = 356$ |

Table 3. Poor man's normalized keys from fence key values and offset-value codes.

| | Lowest code value | Highest code value | Parameter | Formula |
|---|---|---|---|---|
| Low fence keys | 0 | $capacity - 1$ | $run$ | $run$ |
| Current run | $capacity$ | $maxuint/2$ | $ovc$ | $ovc + capacity$ |
| Next run | $maxuint/2 + 1$ | $maxuint - capacity$ | $ovc$ | $ovc + maxuint/2 + 1$ |
| High fence keys | $maxuint - capacity + 1$ | $maxuint$ | $run$ | $\sim run = maxuint - run$ |

Offset-value coding encodes one row's key value relative to another row's key value. Thus, offset-value codes are set after comparisons, specifically for the loser in a comparison, i.e., the row later in the sort order. The offset is the position where the two rows' keys first differ, e.g., a column index. The value is not truly required but often helpful. If it is used, it is the key value at the offset. It could also be the difference to the winner's value at the offset, but this alternative does not offer much benefit.

Table 2 shows, using the same key values as Table 1, an alternative way of calculating offset-value codes. Here, even an ascending sort uses descending offset-value codes. The constant value 1 ensures non-negative offset-value codes, the value 3 indicates the offset or prefix size, the value 100 is the key domain of each column, and the value 44 is the column value after the offset.

In a tree-of-losers priority queue, offset-value coding has each node x coded relative to the local winner, which is also the winner of the entire sub-tree rooted at node x. As key values move towards the root of the priority queue, their offset-value codes may change when they lose in a comparison and are left behind by a winner key value.

For optimal efficiency of comparisons, a poor man's normalized keys combines fence key values ±∞ with offset-value codes. Moreover, during run generation when the priority queue contains key values assigned to two initial runs on temporary storage, the poor man's normalized keys cached in nodes of the priority queue also encodes the current versus the next run.

Table 3 illustrates how fence key values and (ascending) offset-value codes are mapped to poor man's normalized keys. All values are unsigned integers. Poor man's normalized keys range from 0 to $maxuint$, run identifiers from 0 to the capacity of the priority queue, and offset-value codes from 0 to $maxuint/2 - capacity$. With poor man's normalized keys calculated from fence keys and offset-value codes, many or most comparisons in a sort operation or merge step are practically free at the cost of a single integer comparison, i.e., a single hardware instruction.

## 3.5 Grouping and aggregation within sort

The simplest implementation of grouping, aggregation, and duplicate removal first sorts the input and then applies in-stream aggregation. For faster execution, in-stream aggregation while writing runs guarantees that no run contains duplicate key values and, therefore, that no run is larger than the final output [3].

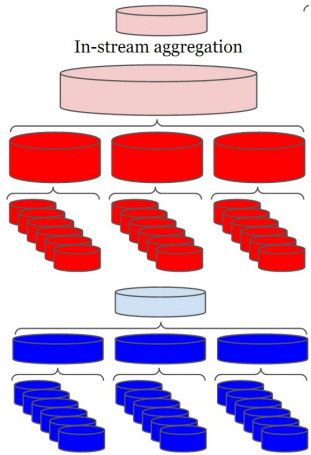

Fig. 2. External merge sort with duplicate removal in runs.

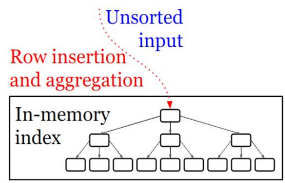

Fig. 3. In-memory aggregation.

Figure 2 illustrates the beneficial effect of duplicate removal within runs. This small example assumes that input and memory sizes force 18 initial runs on temporary storage and that memory and page sizes limit the merge fan-in to 6. On the top, after a traditional external merge sort generates and merges runs, it pipelines the output of the final merge step into an in-stream aggregation operation. The result of the sort is just as large as the unsorted input; only the subsequent in-stream aggregation reduces the data volume. On the bottom, duplicate removal within runs reduces the data volume on temporary storage. Intermediate runs are never larger than the final result, which the final merge step computes.

Better yet, early aggregation detects and eliminates duplicate key values as they arrive from the unsorted input. Early aggregation eschews both quicksort and priority queues. Instead, it uses an ordered in-memory index, e.g., an in-memory b-tree [10]. The index enables immediate discovery of duplicate key values, just like a hash table. If the output size is smaller than the memory size, early aggregation avoids all I/O to spill intermediate data to temporary storage. Figure 3 illustrates early aggregation in this case, i.e., an in-memory b-tree index and no spilling to temporary storage.

If the output size is larger than the memory size, sorted keys must spill into sorted runs on temporary storage. With an in-memory ordered index, a single data structure enables both sorting (instead of a priority queue) and matching (for duplicate removal or aggregation). In many ways, this algorithm is a dual to hybrid hash aggregation, with similar memory requirements, total file size on temporary storage, graceful degradation from an internal to an external algorithm, and possible optimizations for intermediate records that absorb the most input records – one important difference, of course, is the sort order of the output.

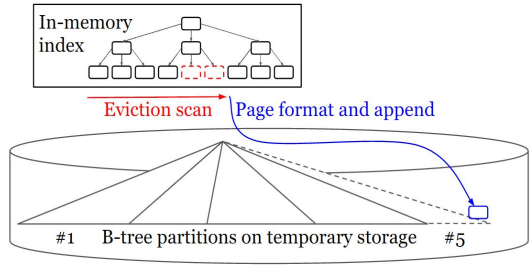

Fig. 4. Run generation using an ordered in-memory index.

Figure 4 illustrates run generation using an ordered in-memory index. An in-memory index serves both aggregation (similar to a hash table in hash aggregation) and run generation (similar to a priority queue in an external merge sort). Traditional run generation can use either read-sort-write cycles, e.g., using quicksort, or replacement selection, e.g., using a priority queue. An in-memory index for run generation enables both read-sort-write cycles and replacement selection. In fact, while data records in a priority queue for replacement selection require a tag (a bit) indicating the current or the next run, data records in an in-memory index do not. There is a scan evicting data records to the current run. Its scan position within the in-memory index separates data records destined for the current run and those for the next run.

## 3.6 Compression in sorted files

If memory space or storage bandwidth are tight, and in particular if CPU bandwidth is plentiful, compression can speed up operations such as external merge sort. Two forms of compression exploit sort order: first, in row-format storage, prefix truncation including offset-value coding removes redundant leading key values; and second, in column-format storage, run-length encoding of sort keys works best for the leading columns in a compound key because the runs of equal values are the longest among all sort key columns. These two techniques are applicable to different storage formats yet they achieve similar compression for leading sort columns.

## 3.7 Summary of related prior work

In summary, there are many techniques, both algorithms and data structures including file formats, that speed up sorting and related activities. The next section focuses on the performance improvements that offset-value coding enables in sorting and in database query execution.

## 4 TECHNIQUES

This section first illustrates what offset-value coding can contribute to efficient sorting and how much it can reduce the number of column comparisons in compound keys. It then introduces new insights and observations, in particular the link between offset-value coding and compression of sorted runs as well as efficient techniques for translating between prefix truncation in row-format runs and run-length encoding of leading columns in column-format runs. This section also introduces offset-value coding in query execution and data processing beyond sorting, e.g., sort-based grouping and duplicate removal, merge join, and more.

Inasmuch as offset-value coding is a dynamic form of poor man's normalized keys, it requires incremental normalization of key values. In other words, it requires key normalization only for a subset of columns even in very large compound keys, often only one or two column values, but it does require the standard normalization considerations, e.g., encoding of *null* values and of descending sort keys.

## 4.1 Merge inputs and output

Let the following small example illustrate the effects of offset-value coding on the required comparisons while merging sorted runs (ascending on columns A, B, C..., J).

Table 4. Merge inputs.

| Row# | A | B | C | D | E | F | G | H | J |
|------|---|---|---|---|---|---|---|---|---|
| 0 | 1 | 1 | 2 | 5 | 1 | 1 | 1 | 1 | 1 |
| 1 | 1 | 1 | 2 | 5 | 1 | 2 | 1 | 1 | 1 |
| 2 | 1 | 1 | 2 | 5 | 1 | 3 | 0 | 0 | 0 |

| Row# | A | B | C | D | E | F | G | H | J |
|------|---|---|---|---|---|---|---|---|---|
| 0 | 1 | 1 | 1 | 1 | 1 | 1 | 1 | 1 | 1 |
| 1 | 1 | 1 | 1 | 2 | 10 | 1 | 1 | 1 | 1 |
| 1 | 1 | 1 | 2 | 5 | 1 | 3 | 0 | 1 | 0 |

| Row# | A | B | C | D | E | F | G | H | J |
|------|---|---|---|---|---|---|---|---|---|
| 0 | 1 | 1 | 2 | 5 | 1 | 3 | 0 | 0 | 1 |
| 1 | 1 | 1 | 2 | 5 | 1 | 3 | 0 | 0 | 2 |
| 2 | 1 | 1 | 2 | 5 | 1 | 3 | 0 | 0 | 3 |
| 3 | 2 | 0 | 0 | 0 | 0 | 0 | 0 | 0 | 0 |
| 4 | 2 | 0 | 0 | 0 | 0 | 0 | 0 | 0 | 1 |
| 5 | 2 | 0 | 1 | 0 | 9 | 9 | 9 | 9 | 9 |
| 6 | 2 | 0 | 2 | 0 | 0 | 0 | 0 | 0 | 0 |

Table 4 shows three merge inputs with multi-column keys; each row has 9 key columns. The shaded column values indicate the first difference from the preceding row; prefix truncation suppresses all columns to the left of a shaded column. The row# is not stored, it is included here only for ease of reference in the discussion below.

Table 5 shows the merge output. Column values are shaded to indicate prefix truncation in the merge output, even if the shading color still indicates the merge input. For example, input row #0 in the green input is output row #2. Prefix truncation can eliminate more leading column values in the output than in the input, here due to the preceding blue row. Similarly, prefix truncation in output row #8 applies to more columns than in the same row within the blue input, where it is row #2. For optimal compression in merge inputs and output, the merge logic must determine not only the sequence of output rows but also their maximal prefix truncation.

## 4.2 Offset-value coding and compression

It turns out that offset-value coding and prefix truncation are tightly linked. When reading input rows and adding them to a tree-of-losers priority queue, the key values shaded in Table 4 become each row's initial offset-value code. Their position within the input row becomes the offset and their normalized column value becomes the value; the combination of this offset and this value is the key value cached in the nodes of a tree-of-losers priority queue for efficient comparisons.

In other words, a merge step for the inputs shown in Table 4 may ignore the column values shared with preceding rows and truncated as common prefixes. Prefix truncation has spent comparison effort on those columns in the earlier steps that produced these merge inputs; the merge step can exploit the information gained in those earlier comparisons if it is encoded as prefix truncation and developed further with offset-value coding.

Table 5. Merge output.

| Row# | A | B | C | D | E | F | G | H | J |
|------|---|---|---|---|---|---|---|---|---|
| 0 | 1 | 1 | 1 | 1 | 1 | 1 | 1 | 1 | 1 |
| 1 | 1 | 1 | 1 | 2 | 10 | 1 | 1 | 1 | 1 |
| 2 | 1 | 1 | 2 | 5 | 1 | 1 | 1 | 1 | 1 |
| 3 | 1 | 1 | 2 | 5 | 1 | 2 | 1 | 1 | 1 |
| 4 | 1 | 1 | 2 | 5 | 1 | 3 | 0 | 0 | 0 |
| 5 | 1 | 1 | 2 | 5 | 1 | 3 | 0 | 0 | 1 |
| 6 | 1 | 1 | 2 | 5 | 1 | 3 | 0 | 0 | 2 |
| 7 | 1 | 1 | 2 | 5 | 1 | 3 | 0 | 0 | 3 |
| 8 | 1 | 1 | 2 | 5 | 1 | 3 | 0 | 1 | 0 |
| 9 | 2 | 0 | 0 | 0 | 0 | 0 | 0 | 0 | 0 |
| 10 | 2 | 0 | 0 | 0 | 0 | 0 | 0 | 0 | 1 |
| 11 | 2 | 0 | 1 | 0 | 9 | 9 | 9 | 9 | 9 |
| 12 | 2 | 0 | 2 | 0 | 0 | 0 | 0 | 0 | 0 |

As the merge logic proceeds, new input rows reach the root position of the tree-of-losers priority queue. In the process, these input rows may change their offset and thus also the cached key value. When a row reaches the root position, its offset at that time is also the size of its prefix truncation in the merge output. This is because an offset increases only in a loser within a comparison of key values, which is required only if there is another row (originating from a different merge input) with equal values in each preceding column.

Table 6. Comparisons of column values.

| Row# | A | B | C | D | E | F | G | H | J |
|------|---|---|---|---|---|---|---|---|---|
| 0 | 1 | 1 | 1 | 1 | 1 | 1 | 1 | 1 | 1 |
| 1 | 1 | 1 | 1 | 2 | 10 | 1 | 1 | 1 | 1 |
| 2 | 1 | 1 | 2 | 5 | 1 | 1 | 1 | 1 | 1 |
| 3 | 1 | 1 | 2 | 5 | 1 | 2 | 1 | 1 | 1 |
| 4 | 1 | 1 | 2 | 5 | 1 | 3 | 0 | 0 | 0 |
| 5 | 1 | 1 | 2 | 5 | 1 | 3 | 0 | 0 | 1 |
| 6 | 1 | 1 | 2 | 5 | 1 | 3 | 0 | 0 | 2 |
| 7 | 1 | 1 | 2 | 5 | 1 | 3 | 0 | 0 | 3 |
| 8 | 1 | 1 | 2 | 5 | 1 | 3 | 0 | 1 | 0 |
| 9 | 2 | 0 | 0 | 0 | 0 | 0 | 0 | 0 | 0 |
| 10 | 2 | 0 | 0 | 0 | 0 | 0 | 0 | 0 | 1 |
| 11 | 2 | 0 | 1 | 0 | 9 | 9 | 9 | 9 | 9 |
| 12 | 2 | 0 | 2 | 0 | 0 | 0 | 0 | 0 | 0 |

Table 6 shows columns in value comparisons. (In Table 4 and Table 5, shading indicates the first column after prefix truncation; here, shading indicates rows and column values that require comparisons beyond their initial offset-value codes.) For each comparison of a pair of column values (in a pair of rows), only one field is shaded, namely in the loser. In other words, the number of shaded column values in Table 6 equals the number of single-column value comparisons. With offset-value coding, merging the three runs in Table 4 to form the output in Table 5 requires only 15 column value comparisons.

Table 7. New prefix truncation in the merge output.

| Row# | A | B | C | D | E | F | G | H | J |
|------|---|---|---|---|---|---|---|---|---|
| 0 | 1 | 1 | 1 | 1 | 1 | 1 | 1 | 1 | 1 |
| 1 | 1 | 1 | 1 | 2 | 10 | 1 | 1 | 1 | 1 |
| 2 | 1 | 1 | 2 | 5 | 1 | 1 | 1 | 1 | 1 |
| 3 | 1 | 1 | 2 | 5 | 1 | 2 | 1 | 1 | 1 |
| 4 | 1 | 1 | 2 | 5 | 1 | 3 | 0 | 0 | 0 |
| 5 | 1 | 1 | 2 | 5 | 1 | 3 | 0 | 0 | 1 |
| 6 | 1 | 1 | 2 | 5 | 1 | 3 | 0 | 0 | 2 |
| 7 | 1 | 1 | 2 | 5 | 1 | 3 | 0 | 0 | 3 |
| 8 | 1 | 1 | 2 | 5 | 1 | 3 | 0 | 1 | 0 |
| 9 | 2 | 0 | 0 | 0 | 0 | 0 | 0 | 0 | 0 |
| 10 | 2 | 0 | 0 | 0 | 0 | 0 | 0 | 0 | 1 |
| 11 | 2 | 0 | 1 | 0 | 9 | 9 | 9 | 9 | 9 |
| 12 | 2 | 0 | 2 | 0 | 0 | 0 | 0 | 0 | 0 |

Table 7 shows the 15 column values compressed (prefix truncated) in the merge output but not in the merge inputs. Note the systematic difference between Table 6 and Table 7. In the three affected rows, prefix truncation in the output includes the column value shaded in Table 4 but not those shaded in Table 5. The number of additional column values in prefix truncation is precisely equal to the number of column value comparisons.

When sorting $N$ rows with $K$ key columns, it is not possible to truncate more than $K \times (N-1)$ column values. Therefore, this sort has no need for more than $K \times (N-1)$ comparisons of column values. In other words, even with $O(N \log_N)$ or more precisely $N \times \log_2(N/\epsilon)$ comparisons of offset-value codes, the comparison effort in terms of column comparisons is linear with the row count and with the column count, or $O(K \times N)$. The case of maximal truncation and maximal comparison effort is an input with $N$ duplicate key values.

## 4.3 Translation between compressed row- and column-formats

For sorted data, the obvious compression techniques are prefix truncation for storage in row-format and run-length encoding for storage in column-format. Both formats permit the same amount of compression due to the sort order. Both formats also permit other compression techniques, e.g., dictionary encoding as well as traditional and order-preserving Huffman compression.

Table 8 shows a table with 9 rows and 5 columns, sorted on all columns and compressed using prefix truncation. Different shading merely indicates separate rows. Only shaded column values are stored explicitly; the column values not shaded are truncated and replaced by the artificial column indicating the size of the truncated prefix. If all rows have logically the same number of columns, a count of the fields in a storage record suffices to indicate the count of columns suppressed due to prefix truncation. In this example, 19 of 45 column values require no storage due to prefix truncation.

Table 9 shows the same rows, columns, and values in column-format, compressed using run-length encoding. Different shading merely indicates separate columns and only shaded column values are stored explicitly. Precisely the same column values are suppressed or stored in Table 8 and Table 9, although the compression information is organized differently. Not surprisingly, the sum of prefix sizes in Table 8 is precisely equal to the sum of run lengths in Table 9.

Table 8. Prefix truncation in a row-format.

| Row# | Prefix | A | B | C | D | E |
|------|--------|---|---|---|---|---|
| 0 | 0 | | | | | |
| 1 | +1 | | | | | |
| 2 | +2 | | | | | |
| 3 | +4 | | | | | |
| 4 | +4 | | | | | |
| 5 | +4 | | | | | |
| 6 | +3 | | | | | |
| 7 | +1 | | | | | |
| 8 | 0 | | | | | |

Table 9. Run-length encoding in a column-format.

| Row# | A | B | C | D | E |
|------|---|---|---|---|---|
| 0 | | | | | |
| 1 | +7 | | | | |
| 2 | | +5 | | | |
| 3 | | | +4 | +3 | |
| 4 | | | | | |
| 5 | | | | | |
| 6 | | | | | |
| 7 | | | | | |
| 8 | | | | | |

Run generation or a merge step produce sorted data one row at a time. If storage in column-format is desired or required, the translation is surprisingly simple. For example, writing row 0 (prefix size 0) adds a value to each column; writing row 1 (prefix size 1) adds a value only to the 4 trailing columns; etc.; writing row 8 first adds a run length (indicating 7 additional values) to column 1 before adding its own value to each column; and similarly for runs in other columns.

It is not required to maintain all run lengths for each row. For example, when adding row 3 (prefix size 4), only the run-length indicator for column D requires maintenance, whereas the run length indicators for columns A-C may remain unchanged. When adding row 6 (prefix size 3), the current run length indicator for column D (value +3) is added to that of column C. Thus, when adding row 7 (prefix size 1), the run-length indicator for column C is 4. Maintenance of the run length indicators resembles longhand addition carrying overflow information from one digit position to the preceding one. Thus, we call this translation from row compression to column compression "with carry."

When a merge step reads its input runs in column-format, it must translate them into rows with prefix truncation. That translation is similarly simple and similarly permits a variant "with carry." Note that merge logic with offset-value coding has no need for the column values without shading in Table 8 and Table 9. From each run in run-length encoding, the merge logic and its comparisons require only the value defining the run. Nevertheless, if all column values are desired for some reason, the translation from column to row-formats remains simple and efficient, with no need for any comparisons of column values.

## 4.4 Prefix truncation in initial runs

Merging runs using tree-of-losers priority queues benefits from offset-value coding, which in turn is optimally initialized by prefix truncation in the merge inputs. Obvious questions include how to create prefix truncation in initial runs and the overhead this introduces.

If run generation relies on read-sort-write cycles, e.g., using quicksort, prefix truncation can be introduced while writing initial runs or while reading them. For compression and low I/O volume, prefix truncation after sorting and before writing seems ideal. An alternative to read-sort-write is replacement selection, e.g., using a priority queue. In fact, if a tree-of-losers priority queue is used, offset-value coding can be used not in merge steps but even in run generation.

When merging M runs with a single row each, each row has an initial offset of zero as appropriate for the first row in its run. Thus, it is easy to determine the appropriate offset-value code for each row. As rows are "merged," their offsets increase in the same way as in any other merge step.

Replacement selection with initial runs larger than memory must compare each row from the unsorted input with the most recently evicted key value. This comparison of new input key value and evicted key value determines the offset-value code of the new input row. If the new key value is the winner in the comparison, i.e., if it is deferred to the next run on temporary storage, the new row is the first row in the next initial run on temporary storage, i.e., the first row in a "run" with only a single row, and thus its offset is zero. Otherwise, the comparison assigns an offset-value code to the loser in the standard way. Run generation by merging M runs continues and exploits the offset-value code of the new key value.

If replacement selection evicts and reads individual rows, variable-size rows introduce fragmentation and thus a workspace management problem. Efficient insertion of new unsorted rows by first-fit or even best-fit incurs 10-25% loss in run size [21]. For example, instead of runs 2× the size of memory, their size might be 1.5-1.8× the size of memory.

## 4.5 Offset-value coding in sort-based query execution

Information about offsets, prefixes, or run lengths is readily available for the output of a sort operation. It can be used to write a compressed sorted file, either in row- or column-format, but it can also be used in a subsequent query operation. In other words, offset-value coding is useful not only within a sort operation but also more generally for sort-based query execution and its specific execution algorithms. For example, if a merge join obtains its two inputs with offset-value codes attached to each row, either from a sort operation or from a compressed file, these offset-value codes can speed up comparisons within the merge join. After all, the core logic of a merge join is a merge operation just like a merge step in external merge sort.

Another operation that primarily merges its input is an order-preserving re-partitioning or exchange operation [13]. Either a single consumer or each consumer receives sorted streams from multiple producers and merges those into a single sorted stream per consumer. If the data packets sent from producers to consumers include offset-value codes, they can be compressed using prefix truncation or run-length encoding and they enable efficient merging with the minimal number of column comparisons.

In addition to these query execution operations that are essentially merge steps, offset-value codes can also benefit other operations that check for equal key values among neighbors in a stream. A row and its key value are equal to the predecessor in a sorted stream if prefix truncation applies to the entire key or, equivalently, if the offset in an offset-value code exceeds the key size.

Several such operations come to mind immediately. First, in-stream aggregation exploits sort order for efficient execution of "group by" queries. Second, in-stream duplicate removal is a special (simple) form of in-stream aggregation. An external merge sort with duplicate removal or even aggregation integrated in its sort and merge logic similarly can find duplicate key values by an offset in an offset-value code exceeding the key size. Third, in-stream segmentation divides a stream into sub-streams defined by equal values in leading sort keys. The first row of a new sub-stream is easily found by an offset lower than the size of the key that defines segments.

Segmented query execution can turn a single large operation into multiple smaller operations. A typical example is extending the list of sort keys. For example, sorting on key list (A, B, C, D) can exploit an input sorted on (A, B) or (A, B, X, Y). Instead of a single large sort operation, likely using an external merge sort, segmented execution sorts on (C, D) each sub-stream of rows with equal values in (A, B), likely using an in-memory sort for each segment. In the first row within a segment, the offset in offset-value codes is initialized to the size of the segmentation key, not to zero as in most other sort operations. Operations other than sorting can benefit from segmented execution, e.g., grouping and duplicate removal. Segmentation does not strictly require sorting, only grouping of equal key values.

### 4.6 Summary of techniques

In summary of new techniques, external merge sort can exploit offset-value codes not only for faster comparisons, in fact with linear effort in terms of column comparisons, but also to compress run files on temporary storage, in the forms of both prefix truncation in row-formats and run-length encoding in column-formats, and to speed up query execution algorithms on sorted data, from merge join to duplicate removal, grouped aggregation, and segmented execution.

## 5 PERFORMANCE EVALUATION

The value of experiments is to validate or refute hypotheses created from traditional or new understanding of a subject matter. The following hypotheses seem worth testing for a deeper understanding of sorting, sort-based index creation, index maintenance, and query processing.

### 5.1 Hypotheses

Among the following hypotheses, the first two are included merely to confirm that tree-of-losers priority queues are a good foundation for efficient sorting. The remaining hypotheses test the value of new techniques introduced here.

(1) In practical implementations of run generation, tree-of-losers priority queues require fewer row comparisons than quicksort, both in the average case and in the worst case.
(2) When merging runs, tree-of-losers priority queues require fewer row comparisons than standard priority queues.
(3) Comparisons of offset-value codes are just as fast as comparisons of poor man's normalized keys (i.e., inline integer comparisons, practically free if folded in with fence keys).
(4) When it is effective, offset-value coding reduces comparisons of column values by a factor proportional to the length of relevant keys.
(5) When sorting on hash values, offset-value coding is not effective but its overhead is negligible.
(6) Offset-value coding enables efficient detection and management of duplicate key values, e.g., when merging runs for duplicate removal or for grouped aggregation.
(7) Offset-value coding speeds up compression of intermediate runs and of the final output, both prefix truncation in row-formats and run-length encoding in column-formats.

(8) Offset-value coding can speed up not only external merge sort but also consumers of its output, i.e., of sorted streams with prefix truncation, e.g., in-stream aggregation and merge join but also segmentation and order-preserving exchange (merging shuffle).

The experiments below are designed specifically to support or refute these hypotheses.

## 5.2 Testing environment

In order to focus on the sort operation, priority queues, and offset-value coding, all experiments use a single node and a single execution thread. The hardware is a contemporary workstation; more detailed specifications are omitted on purpose. Each experiments starts with a warm cache, i.e., input data pre-fetched into memory. The measurements below come from Google's public benchmark library [12]. Test data are synthetic yet similar to the actual data in our web analysis workload running in production day in, day out. Thus, there are many rows with many key columns. Each key column is an 8-byte integer but with only very few actual values.

## 5.3 Algorithm implementations

This section briefly summarizes extensions of Google's F1 Query engine [25, 27] that employ offset-value coding.

F1 Query is a federated query processing platform that executes SQL queries against data stored in different storage systems at Google such as Spanner [8] and BigTable [5]. Written in C++, its execution kernel is column-oriented. The sort operator in F1 Query has two major components: a run generator that produces initial sorted runs from unsorted input and a run merger. Traditionally, the two components used an interpreted row comparator that handles all supported data types, null values, and sort direction and collation. As a result, comparing rows with many key columns was substantially more expensive than, for example, an integer comparison compiled with the sort operator.

The tree-of-losers priority queue extension adds offset-value coding. Each entry in the tree encodes offset and value bits in an unsigned 64-bit integer offset value code (OVC). The tree's API is extended so that callers can provide an OVC for each new data item and obtain an OVC for each item in the merge output.

The modified sort operator takes advantage of the tree-of-losers priority queue. First, instead of standard quicksort, the run generation component uses a tree-of-losers priority queue to sort N rows by modeling each record as a separate run. Thus, sorting N rows is like merging N one-row runs with the tree-of-losers priority queue. As importantly, this new algorithm computes the OVC for each row in the sorted run as a by-product. Second, the run merger lets duplicate key values bypass the merging queue. Specifically, the run merger inspects the OVC of the next row from the winner's input, and if this next key value has an offset equal to the number of key columns, it copies the row directly to the output buffer.

F1 Query takes advantage of OVCs from sorted intermediate results (e.g., the output of the sort operator) to further speed up data processing. Specifically, F1 Query supports carrying OVCs between operators whenever possible. A classic example is carrying OVCs from the sort to a subsequent in-stream aggregation, where OVCs are used to detect duplicate grouping keys and group boundaries. Other examples include carrying OVCs in a distributed sort from local sort to merging exchange, from sort to merge join, and from merge join to in-stream aggregation. Query optimization in F1 Query introduces an artificial column representing OVCs for any order-producing physical operator in a query plan. F1 currently uses OVCs to speed up data processing in in-sort aggregation [10], segmented execution, in-stream aggregation, merge join, and analytic functions. The sorting techniques described in this paper are also implemented and deployed for
Napa [1].

## 5.4 Measurements

In this section, we present experimental results that test the hypotheses of Section 5.1. Even without
a traditional rigorous analysis with support levels and confidence intervals, the goal is to confirm
or refute these hypotheses.

*5.4.1 Counts of row comparisons in run generation.* Hypothesis 1 claims that in practical imple-
mentations of run generation, tree-of-losers priority queues require fewer row comparisons than
quicksort, both in the average case and in the worst case.

Table 10. Counts of row comparisons in run generation.

| Row count | Quicksort | Tree of losers | Lower bound | Factor |
|----------:|----------:|---------------:|------------:|-------:|
| 1,000 | 11,696 | 8,722 | 8,525.8 | 1.023 |
| 10,000 | 160,859 | 120,949 | 118,477.1 | 1.021 |
| 100,000 | 2,020,269 | 1,542,713 | 1,516,964.0 | 1.017 |
| 1,000,000 | 24,133,548 | 18,687,584 | 18,491,568.6 | 1.011 |

Table 10, in a test of hypothesis 1, shows the count of row comparisons for sorting 1,000 to
1,000,000 rows with key values in uniform distributions. The comparison counts for quicksort
and tree-of-losers priority queue are measured; the lower bound is calculated using $\log_2(N!) \approx
N \times \log_2(N/\epsilon)$. The factor indicates how close the tree-of-losers priority queue is to the lower
bound.

It is evident that a tree-of-losers priority queue requires fewer row comparisons than quicksort,
close to the lower bound of required comparisons. In a tree-of-losers priority queue, the number of
comparisons is practically constant for any distribution of key values, whereas quicksort requires
$O(N^2)$ comparisons in the worst case.

*5.4.2 Counts of row comparisons in merge steps.* Hypothesis 2 claims that while merging runs,
tree-of-losers priority queues require fewer row comparisons than standard priority queues.

Table 11. Counts of row comparisons in merging.

| Row count per run | std::priority_queue | Tree of losers | Factor |
|------------------:|--------------------:|---------------:|-------:|
| 1,000 | 39,427 | 23,981 | 1.644 |
| 10,000 | 394,286 | 239,983 | 1.643 |
| 100,000 | 3,940,481 | 2,399,985 | 1.642 |
| 1,000,000 | 39,417,524 | 23,999,992 | 1.642 |

Table 11, in a test of hypothesis 2, shows the count of row comparisons while merging 8 runs,
each with the given row count. The factor indicates the relative comparison count for the standard
priority queue and a tree-of-losers priority queue. Whereas the standard library implementation of
a priority queue requires both root-to-leaf and leaf-to-root passes, a tree-of-losers priority queue
requires only one leaf-to-root pass per input row and each pass requires $\log_2(F)$ comparisons.
With a merge fan-in of F=8, 1,000 rows per run times 8 runs times $\log_2(8) = 3$ comparisons per
row gives 24,000 row comparisons or a few less during initialization and at end-of-input. Thus,
not only do tree-of-losers priority queue require fewer comparisons than standard priority queue
implementation, their count of comparisons is very close to the minimum.

Table 12.  CPU effort for comparisons [$\mu s$].

| Common key prefix | Row count = 1K | | Row count = 10K | | Row count = 100K | | Row count = 1M | |
|---|---|---|---|---|---|---|---|---|
| | PNMK | OVC | PMNK | OVC | PMNK | OVC | PMNK | OVC |
| 0 | 111 | 175 | 1,318 | 1,997 | 10,657 | 26,840 | 119,739 | 421,528 |
| 2 | 105 | 178 | 1,031 | 2,023 | 12,815 | 27,771 | 219,300 | 428,910 |
| 4 | 369 | 176 | 4,828 | 2,082 | 63,557 | 27,560 | 1,111,267 | 443,536 |
| 6 | 523 | 185 | 5,773 | 2,114 | 79,004 | 28,364 | 1,262,471 | 436,634 |
| 8 | 509 | 183 | 6,928 | 2,154 | 91,784 | 28,505 | 1,404,161 | 459,768 |

Table 13.  Counts of column value comparisons.

| Common key prefix | Row count = 1K | | Row count = 10K | | Row count = 100K | | Row count = 1M | |
|---|---|---|---|---|---|---|---|---|
| | NoOVC | OVC | NoOVC | OVC | NoOVC | OVC | NoOVC | OVC |
| 0 | 5,132 | 999 | 68,752 | 9,999 | 902,832 | 99,999 | 10,238,080 | 999,999 |
| 2 | 15,396 | 2,997 | 206,256 | 29,997 | 2,708,496 | 299,997 | 30,714,240 | 2,999,997 |
| 4 | 25,660 | 4,995 | 343,760 | 49,995 | 4,514,160 | 499,995 | 51,190,400 | 4,999,995 |
| 6 | 35,924 | 6,993 | 481,264 | 69,993 | 6,319,824 | 699,993 | 71,666,560 | 6,999,993 |
| 8 | 46,188 | 8,991 | 618,768 | 89,991 | 8,125,488 | 899,991 | 92,142,720 | 8,999,991 |

*5.4.3 CPU effort for comparisons.* Hypothesis 3 claims that comparisons of offset-value codes are just as fast as comparisons of poor man's normalized keys (i.e., inline integer comparisons) and practically free if folded in with fence keys.

Table 12, in a test of hypothesis 3, shows the CPU effort [in $\mu s$] for comparisons while sorting, e.g., for run generation, with either a poor man's normalized key (PMNK) or offset-value coding (OVC). The common key prefix indicates the number of leading key columns that are constant, i.e., that contribute only overhead to each comparison.

With no common key prefix or a short common key prefix, the poor man's normalized key can capture the relevant column and thus all or most comparisons are very fast. In those cases, the dynamic nature of offset-value coding adds a little overhead. With a large common key prefix, a poor man's normalized key becomes practically useless. In contrast, offset-value codes are much faster for large keys and they show consistent performance for small and large keys. In other words, describing offset-value codes as dynamic poor man's normalized keys seems fully justified.

*5.4.4 Counts of column value comparisons.* Hypothesis 4 claims that when it is effective, offset-value coding reduces comparisons of column values by a factor proportional to the length of relevant keys.

Table 13, in a test of hypothesis 4, shows counts of column value comparisons while sorting, e.g., for run generation, with and without offset-value codes. Note that these are column value comparisons, not row comparisons. In each measurement, the same column decides all comparisons, whereas all other (prefix) columns hold the same value in all rows.

With offset-value coding, a few column comparisons advance the comparison logic to the deciding column, whereas the traditional sort and comparison logic compares prefix columns repeatedly and wastefully. The OVC columns in Table 13 show comparison counts effectively equal to the product of row count and (key) column count; each column value comparison advances an offset such that one such column is not compared again. An analysis within each row of Table 13 illustrates how the number of column value comparisons is linear in the row count (or the product of row count and column count) even if the total number of row comparisons reflects the $O(N \times \log(N))$ complexity of comparison-based sorting. In other words, most comparisons are decided by offset-value codes, not by comparisons of column values.

Table 14. CPU effort for comparisons of hash values [$\mu s$]

| Row count | Neither | Poor man's normalized key | Offset-value coding |
|---|---|---|---|
| 1,000 | 245 | 135 | 134 |
| 10,000 | 3,448 | 1,575 | 1,521 |
| 100,000 | 49,101 | 20,076 | 18,471 |
| 1,000,000 | 814,573 | 333,448 | 264,969 |

Table 15. CPU effort for merging with duplicate key values [$\mu s$].

| Copies per distinct key | 8 key columns | | |
|---|---|---|---|
| | Full-key comparison | Offset-value coding | Factor |
| 1 | 224,745 | 103,679 | 2.168 |
| 10 | 189,511 | 86,635 | 2.187 |
| 100 | 130,724 | 56,662 | 2.307 |
| 1,000 | 124,446 | 43,562 | 2.857 |
| 10,000 | 112,981 | 41,019 | 2.754 |

*5.4.5    CPU effort for OVC comparisons.* Hypothesis 5 claims that when sorting on hash values, offset-value coding is not effective but its overhead is negligible.

Table 14, in a test of hypothesis 5, shows the CPU effort [in $\mu s$] for sorting (run generation) with poor man's normalized keys, offset-value codes, and neither of these two techniques. In all cases, column 0 decides all comparisons, i.e., it is the best case for poor man's normalized keys.

Poor man's normalized keys are much more efficient than full comparisons, because a poor man's normalized key is computed once per row whereupon all comparisons require only a single (inline) integer instruction, whereas a full comparison always requires a function invocation as well as a loop and some limited interpretation of per-column metadata. Offset-value coding seems slightly more efficient than poor man's normalized keys; we have not determined the reason for this difference despite earnest efforts with profiling etc. The large difference to the traditional comparison logic is due to inline comparisons of offset-value codes as single integers.

*5.4.6    CPU effort for duplicate detection.* Hypothesis 6 claims that offset-value coding enables efficient detection and management of duplicate key values, e.g., when merging runs for duplicate removal or for grouped aggregation.

Table 15, in a test of hypothesis 6, shows the benefit of offset-value coding compared to traditional full comparisons in merging 8 runs with 100,000 rows each. Table 15 reports only the effort for the priority queue and its comparisons. The right-most column indicates the relative performance of merging with full key comparisons or offset-value codes.

With 100 or even 1,000 duplicate key values scattered in 8 merge inputs, each merge input contains about 12 or even 125 copies of each distinct row. With 10 duplicate key values in 8 merge inputs, at least 20% of all key values are duplicates of their predecessors in their respective input runs. In all those cases, many input rows can bypass the priority queue and its comparisons. With full comparisons, each input row requires one full comparison to determine whether it is a duplicate key value and can therefore bypass the priority queue; with offset-value coding, the offset alone can decide this question simply and efficiently. Therefore, the performance advantage of offset-value coding grows with an increasing number of duplicate key values in the merge inputs.

Table 16. Effort for compression by truncating shared prefixes [$\mu s$].

| #Key columns | Full-key comparison | Offset-value coding | Compression ratio |
|---:|---:|---:|---:|
| 2 | 148,354 | 118,853 | 4 |
| 4 | 108,257 | 67,014 | 19 |
| 6 | 76,493 | 9,656 | 1,353 |
| 8 | 73,900 | 8,618 | 102,040 |

Table 17. Effort for compression by run-length encoding [$\mu s$].

| #Key columns | Full | OVC | Compression ratio |
|---:|---:|---:|---:|
| 2 | 150,943 | 63,048 | 4 |
| 4 | 141,942 | 36,745 | 19 |
| 6 | 137,442 | 6,863 | 1,353 |
| 8 | 129,705 | 6,347 | 102,040 |

### 5.4.7 Compression with offset-value coding.
Hypothesis 7 claims that offset-value coding speeds up compression of intermediate runs and of the final output, both prefix truncation in row-formats and run-length encoding in column-formats.

Table 16, in a test of hypothesis 7, shows the benefit of offset-value codes during compression of an intermediate result, e.g., a run on temporary storage during an external merge sort. The chosen compression technique truncates leading columns with values equal to those in the preceding row. In all cases, 1,000,000 input rows have 10 columns with some leading key columns equal in all rows and the remaining column values with random integers between 1 and 9. The compression ratios reflect truncation of leading key columns. For example, with 8 leading columns equal in all rows and only 2 columns with values, the output contains only $9^2 = 81$ distinct rows. Thus, 9 rows have 8 columns truncated in that row of Table 16, 72 rows have 9 columns truncated, and the remaining 999,919 rows have all key columns truncated.

If each row requires a full row comparison with its predecessor, compression costs are substantial. If offset-value codes indicate the size of the prefix that can be truncated, no column value comparisons are required and compression is very fast but just as effective.

Table 17, in a second test of hypothesis 7, shows the benefit of offset-value codes during compression by run-length encoding, which would be suitable in columnar storage. As only leading key columns are compressed in this experiment, the compression ratios are precisely the same as in Table 16.

Traditional run-length encoding requires comparisons for all column values; if a sorted stream is available with offset-value codes, no column comparisons are required and compression is very fast yet equally effective.

### 5.4.8 Query speed-up using offset-value coding.
Hypothesis 8 claims that offset-value coding can speed up not only external merge sort but also consumers of its output, i.e., of sorted streams with prefix truncation, e.g., in-stream aggregation and merge join but also segmentation and order-preserving (merging) exchange.

Table 18, in a test of hypothesis 8, shows the benefit of offset-value coding across multiple operations in a query execution plan. In this example, a sort operation exploits offset-value codes

Table 18. Segmentation effort [$\mu s$].

| % of rows flagged | 2 key columns | | 4 key columns | | 6 key columns | | 8 key columns | |
|---|---|---|---|---|---|---|---|---|
| | Full | OVC | Full | OVC | Full | OVC | Full | OVC |
| 100.00 | 61,204 | 53,218 | 61,789 | 50,182 | 62,264 | 50,462 | 62,618 | 56,294 |
| 10.00 | 65,477 | 49,569 | 69,337 | 50,630 | 74,783 | 50,200 | 78,805 | 50,550 |
| 1.00 | 67,140 | 49,936 | 70,343 | 55,930 | 77,238 | 53,325 | 80,945 | 50,303 |
| 0.10 | 69,183 | 49,609 | 72,484 | 53,531 | 82,923 | 50,275 | 92,597 | 50,780 |
| 0.01 | 70,107 | 50,936 | 77,177 | 51,003 | 89,701 | 51,269 | 93,385 | 49,731 |

Table 19. In-stream aggregation effort [$\mu s$].

| Input/Output ratio | 2 key columns | | 4 key columns | | 6 key columns | | 8 key columns | |
|---|---|---|---|---|---|---|---|---|
| | Full | OVC | Full | OVC | Full | OVC | Full | OVC |
| 1 | 139,058 | 111,144 | 150,616 | 109,399 | 159,820 | 111,273 | 165,968 | 111,715 |
| 10 | 136,513 | 61,569 | 138,817 | 63,622 | 141,857 | 63,227 | 147,565 | 61,980 |
| 100 | 131,373 | 51,496 | 134,974 | 51,599 | 139,864 | 51,310 | 146,513 | 51,898 |
| 1,000 | 131,513 | 51,768 | 138,138 | 51,973 | 143,808 | 50,889 | 147,226 | 50,780 |
| 10,000 | 127,463 | 50,936 | 130,950 | 51,003 | 138,880 | 51,269 | 144,089 | 51,964 |

and exports them to a segmentation operation, i.e., an operation to flag rows that start a new segment in segmented query execution. All 1,000,000 rows are passed through by reference, without copying. Rows within a segment share the same prefix; thus, in a sorted stream with offset-value codes, an offset shorter than the desired prefix indicates a segment boundary.

Table 18 shows the effort within the segmentation operation; it is obvious that offset-value codes permit fast detection of segment boundaries irrespective of the number of key columns, whereas full key comparisons suffer from many key columns. Few flagged rows mean many equal key values, which require expensive comparisons of all columns with full comparisons but do not affect the cost of duplicate key detection using offset-value coding.

Table 19, in a second test of hypothesis 8, shows the effort in an in-stream aggregation operation as it exploits the offset-value codes from a preceding sort operation for fast detection of group boundaries. All queries of the type "select count (distinct . . . ) . . . group by . . . " require such a two-step process. Within the sort output, testing the offset against the count of grouping columns is much faster than full comparisons of multiple key columns. Thus, not only sorting but also many other query execution operations can benefit from offset-value codes.

## 5.5 Summary of performance results

In summary of our experimental evaluation of offset-value coding in external merge sort, the experiments specifically designed to support or refute our hypotheses in Section 5.1 mostly support them, thus demonstrating the value of offset-value coding for sorting and more broadly for sort-based database query processing.

## 6 SUMMARY AND CONCLUSIONS

In summary, both offset-value coding and tree-of-losers priority queues are simple, effective, and efficient. The strength of the mutual benefit between file formats (prefix truncation, run-length encoding) and data processing (comparisons, sorting, merge join, etc.) recommends sort-based query execution, ideally with hardware support.

While prefix truncation, run-length encoding, offset-value coding, and tree-of-losers priority queues have all been known, the present paper introduces

(1) the link between compression by prefix truncation in sorted input runs and offsets for key values entering a tree-of-losers priority queue;

(2) the link between prefix truncation in the merge output and the offset derived along the path from a leaf to the root within a tree-of-losers priority queue;

(3) prefix computation as part of run generation from unsorted input rows by near-optimal in-memory sorting using tree-of-losers priority queues and offset-value coding;

(4) sorting with linear effort in terms of column comparisons, or sorting N rows with K key columns with $O(N \times \log(N))$ row comparisons but only $K \times N$ comparisons of column values in the worst case;

(5) efficient bi-directional translation (without comparisons of column values) between a row-format with prefix truncation and a column-format with run-length encoding;

(6) offset-value coding in consumers, e.g., merge join, in-stream duplicate removal, and in-stream aggregation; and

(7) offset-value coding in segmenting sorted inputs as well as offset-value coding in sorting each segment.

These techniques combine to offer an external merge sort with

(1) its number of row comparisons near the theoretical lower bound;

(2) many of those comparisons at the cost of comparing integers;

(3) the number of column value comparisons linear in the row count and in the column count of the key, i.e., without a $log(N)$ multiplier;

(4) maximal compression by prefix truncation or run-length encoding in runs on temporary storage and in the final sort output;

(5) compression with no column value comparisons beyond those of the sort logic; and

(6) means to speed up consumers of sorted data by exploiting comparisons and prefix truncation performed within the sort.

In conclusion, exploiting the tight relationship between offsets in comparisons, prefix truncation, and run-length encoding of leading columns leads to both faster sorting and faster compression as well as to faster query execution in sort-based algorithms such as in-stream aggregation and merge join. Thus, their combination and advantages suggest renewed interest in sort-based query execution.